\documentclass[eqsecnum,preprint,showpacs,amsmath,aps,nofootinbib]
  {revtex4}  
\usepackage{epsfig}
\usepackage{amsmath,amssymb}
\usepackage[dvips]{color}
\usepackage[latin1]{inputenc}
\usepackage{graphicx}

\allowdisplaybreaks[4]

\makeatletter
\renewcommand*{\@fnsymbol}[1]{\ensuremath{\ifcase#1\or *\or \dagger\or
    \ddagger\or 
   \mathsection\or **\or \dagger\dagger
   \or \ddagger\ddagger \else\@ctrerr\fi}}
\makeatother

\begin{document}
\title{Towards obtaining Green functions for a Casimir cavity in de Sitter spacetime}

\author{Giampiero Esposito}
\email[E-mail: ]{gesposit@na.infn.it}
\affiliation{Istituto Nazionale di Fisica Nucleare, Sezione di
Napoli, Complesso Universitario di Monte S. Angelo, 
Via Cintia Edificio 6, 80126 Napoli, Italy}

\author{George M. Napolitano}
\email[E-mail: ]{grm.nap@gmail.com}
\affiliation{Centre for Mathematical Sciences, Lund University, 
22100 Lund, Sweden}

\date{\today}

\begin{abstract}
Recent work in the literature has studied rigid Casimir cavities in
a weak gravitational field, or in de Sitter spacetime, or yet other
spacetime models. The present review paper studies the difficult problem of
direct evaluation of scalar Green functions for a Casimir-type apparatus
in de Sitter spacetime. Working to first order in the small parameter
of the problem, i.e. twice the gravity acceleration times the plates' 
separation divided by the speed of light in vacuum, suitable coordinates
are considered for which the differential equations obeyed by the zeroth-
and first-order Green functions can be solved in terms of special 
functions. This result can be used, in turn, to obtain, via the 
point-split method, the regularized and renormalized energy-momentum
tensor both in the scalar case and in the physically more relevant
electromagnetic case.
\end{abstract}

\pacs{03.70.+k, 04.60.Ds}

\maketitle

\section{Introduction}

It is frequently the case, in theoretical physics, that an abstract mathematical framework finds application to the
expression of laws which were previously unknown or evaluation of effects that was otherwise requiring too long a
time. For example, pseudo-Riemannian geometry and tensor calculus were precisely the tool that Einstein needed to
formulate his geometric theory of the gravitational field \cite{Einstein}, and the wave equation describing massless 
spin-${\frac{1}{2}}$ fields was discovered by Weyl well before any phenomenological evidence for the existence of these
particular spinor fields in nature. As further examples, the duality symmetry possessed by vacuum Maxwell theory
was later applied to shed new light on all conceivable string theories \cite{Witten}. On the side of applications, the theory of 
ordinary differential equations provided, as an example among the many, the appropriate framework for understanding
a peculiar link between axial and polar perturbations of a Schwarzschild black hole \cite{Chan1983}.

In recent years, a lot of encouraging theoretical progress has been made on the energy-momentum description of a Casimir
apparatus in a weak gravitational field \cite{Bimo06,Espo08,Napo08}, and 
there are now very strong theoretical reasons to believe that Casimir energy obeys the
modern version of the equivalence principle and hence gravitates \cite{Call02,Full07,Milt08,Shaj08,Chen12,Milt14}. 
For our purposes, it is of basic importance
that a regularized and renormalized energy-momentum tensor $\langle T_{\mu \nu} \rangle$ has been evaluated for a
Casimir apparatus in a weak gravitational field. On denoting by $a$ the plates' separation, by $g$ the gravity
acceleration and by $c$ the speed of light in vacuum one can consider the small parameter
\begin{equation}
\varepsilon \equiv {\frac{2ga}{c^2}}
\end{equation}
and compute, to first order in $\varepsilon$, all components of $\langle T_{\mu \nu} \rangle$ by a careful and patient
application of the point-split method \cite{ChriA, ChriB}, checking also that covariant conservation of 
$\langle T_{\mu \nu} \rangle$ is fulfilled and that quantum Ward identities, relating photon and ghost Green functions,
are satisfied \cite{Bimo06}. A simpler check is also available, i.e., the evaluation of $\langle T_{\mu \nu} \rangle$
for a scalar type Casimir apparatus, for which Dirichlet or Neumann boundary conditions on the plates are imposed
\cite{Espo08,Napo08}, rather than the mixed boundary conditions appropriate for gauge fields \cite{Espo1997, Bimo08}. 

In our short review we shall also find it useful to recall that,
ever since DeWitt made extensive use of the Hadamard-Ruse-Synge world function
\cite{Hadamard,Ruse,Synge} for the evaluation of Green functions in curved
spacetime, with application to radiation damping
in a gravitational field \cite{Brehme},
quantum field theory in curved spacetime \cite{DeWitt75} and full
quantum gravity \cite{DeWitt60}, the topic has attracted a lot of interest
because it is well suited for physical applications, in particular for
obtaining the regularized and renormalized energy-momentum tensor of
quantum fields that are coupled to a classical gravitational background. 

As we said before, in \cite{Bimo06, Espo08, Napo08}, such techniques have been used 
to establish on firm ground the physical prediction \cite{Call02} that a
Casimir apparatus, when put in a weak gravitational field, will experience
a very tiny push in the upwards direction, behaving therefore {\it as if
it were equivalent to an experimental device of negative mass}.
Although the resulting force is so small that only a significant progress
in signal-modulation techniques would make it testable \cite{Call02},
the effect is, conceptually, of extreme interest, and physicists should
have learned, by now, how important can {\it gedanken experiments} 
turn out to be.
 
Later on, the work in \cite{Chen12}, motivated by the recent
discovery that we live in a universe undergoing accelerated expansion
\cite{Perl99}, studied vacuum fluctuation forces in de Sitter spacetime.
What was lacking therein, however, was a systematic perturbative
evaluation of scalar and electromagnetic Green functions from the
complicated differential equations that can be used to define them
in the first place, when supplemented by suitable boundary conditions.

The plan of our review paper is therefore as follows. Section 2 writes
the partial differential equations for a scalar Green function
in closed slicing coordinate system. Section 3 considers 
the integral representation for Green functions of partial
differential operators on curved Riemannian manifolds, which is
an useful byproduct of the symbolic calculus for more general,
pseudodifferential operators on Riemannian manifolds.
Concluding remarks and open problems are presented in section 4.

\section{Partial differential equations for the scalar 
Green function in closed slicing coordinate system}

It is well known that de Sitter spacetime can be viewed as an hyperboloid
embedded in flat five-dimensional Minkowski spacetime \cite{Hawk73} 
with coordinates $x^{0},x^{1},x^{2},x^{3},x^{4}$. That is, it is defined as the hypersurface with equation
\begin{equation}
-(x^0)^2 + (x^1)^2 + (x^2)^2 + (x^3)^2 + (x^4)^2 = L^2,
\label{(2.1)}
\end{equation}
with $L>0$, embedded in the space $\mathbb{R}^5$ with metric
\begin{equation}
{\rm d}s^2 = -({\rm d}x^0)^2 + ({\rm d}x^1)^2 + ({\rm d}x^2)^2 
+ ({\rm d}x^3)^2 + ({\rm d}x^4)^2.
\label{eq:metr_flat}
\end{equation}
The metric induced on the hyperboloid by the ambient metric (\ref{eq:metr_flat}) can be written as
\begin{equation}
h_{\mu\nu} = \eta_{\mu\nu} + \frac{x_\mu x_\nu}{(L^2 - \eta_{\alpha\beta} x^\alpha x^\beta)},
\label{(2.3)}
\end{equation}
where $\mu,\nu = 0,1,2,3$ and $\eta_{\mu\nu} = \text{diag}(-1,1,1,1)$ is the flat four-dimensional Minkowski metric. 

By following \cite{Chen12}, the joint effect of de Sitter and weak gravitational field can then be described by the metric tensor
\begin{equation}
\begin{split}
g_{\mu\nu} & = h_{\mu\nu} - \frac{\varepsilon x^3}{a} \, \delta_{\mu 0} \delta_{\nu 0} \\
& = \eta_{\mu\nu} + \frac{x_\mu x_\nu}{(L^2 - \eta_{\alpha\beta} x^\alpha x^\beta)} 
- \frac{\varepsilon x^3}{a} \, \delta_{\mu 0} \delta_{\nu 0}, 
\end{split}
\label{eq:metr_ds_gf}
\end{equation}
where, following Refs. \cite{Bimo06, Espo08, Napo08}, 
$\varepsilon \equiv \frac{2 g a}{c^2}$, $a$ being the plates' separation
of the Casimir apparatus, while $g$ is the gravity acceleration and
$c$ the speed of light in vacuum.

At this stage, we would like to remark that in this work we follow a different approach than used 
in \cite{Chen12}. Indeed, while in \cite{Chen12} the author considers a second-order approximation, near $x = 0$, 
of the metric (\ref{eq:metr_ds_gf}), that is  
\begin{equation}
g^{(2)}_{\mu\nu} = \eta_{\mu\nu} + \frac{x_\mu x_\nu}{L^2} - \frac{\varepsilon x^3}{a} \, \delta_{\mu 0} \delta_{\nu 0},
\label{eq:metr_ds_gf_approx}
\end{equation}
here we prefer to keep the metric intact in all the following calculations. This choice is motivated by the fact that, 
in the derivation of the Green function equations, we will obtain a differential equation that, despite its apparent difficulty, 
is known to be be exactly solvable. Clearly, this would not be the case if we considered 
the approximation (\ref{eq:metr_ds_gf_approx}), instead.

Now, by using the so-called closed-slicing coordinate system $\omega \equiv (t,\theta_1,\theta_2,\theta_3)$ defined by
\begin{gather}
x^{0} \equiv L \sinh(t/L),\\
x^{1} \equiv L \cosh(t/L)\cos \theta_{1},\\
x^{2} \equiv L \cosh(t/L) \sin \theta_{1} \cos \theta_{2},\\
x^{3} \equiv L \cosh(t/L) \sin \theta_{1} \sin \theta_{2} \cos \theta_3,\\
x^{4} \equiv L \cosh(t/L) \sin \theta_{1} \sin \theta_{2} \sin \theta_{3},
\end{gather}
the metric in eq. (\ref{eq:metr_ds_gf}) becomes diagonal and has the form
\begin{equation}
\begin{split}
g_{\mu \nu} & =  {\rm diag} \left(-1, {L^{2} \cosh^{2}(t/L)},
{L^{2} \cosh^{2}(t/L) \sin^{2}\theta_{1} },
{L^{2} \cosh^{2}(t/L) \sin^{2}\theta_{1} \sin^{2}\theta_{2}} \right) \\
 & -  \varepsilon \frac{L}{a} \, \cosh^{3}(t/L) \sin \theta_{1} 
\sin \theta_{2} \cos \theta_{3} 
{\rm diag}(1,0,0,0).
\end{split}
\end{equation}
With this choice of coordinates, the scalar curvature is given by
\begin{equation}
R={12 \over L^{2}}-{3 \varepsilon \over 2aL}\cosh(t/L)
(7 \cosh(2t/L)-5) \sin \theta_{1} \sin \theta_{2} \cos \theta_{3}.
\label{(2.12)}
\end{equation}
We collect in appendix A the complete list of Christoffel symbols.

The partial differential equation for the scalar Green function
$G(\omega,\omega')$, in case of arbitrary coupling to the gravitational field,
with parameter $\xi$, reads as
\begin{equation}
(\Box -\xi R) \, G(\omega,\omega')= - \frac{\delta(\omega-\omega')}{\sqrt{-\det g_{\mu\nu}}}.
\label{(2.13)}
\end{equation}
As a next step, following the technique of Refs. 
\cite{Bimo06, Espo08, Napo08}, we consider the asymptotic expansion
of differential operator, scalar curvature and Green function in powers 
of the dimensionless parameter $\varepsilon$ in (1.1), i.e.
\begin{equation}
\Box \sim \Box^{0}+\varepsilon \Box^{1}+{\rm O}(\varepsilon^{2}),
\label{(2.14)}
\end{equation}
\begin{equation}
R \sim R^{0}+\varepsilon R^{1}+{\rm O}(\varepsilon^{2}),
\label{(2.15)}
\end{equation}
\begin{equation}
G(\omega,\omega') \sim G^{0}(\omega,\omega')+\varepsilon G^{1}(\omega,\omega')
+{\rm O}(\varepsilon^{2}).
\label{(2.16)}
\end{equation}
Working to first order in $\varepsilon$, the differential equation
for the full Green function can be therefore split into a pair
of partial differential equations for the zeroth- and first-order terms,
respectively, i.e.
\begin{equation}
\Bigr(\Box^{0}-\xi R^{0}\Bigr)G^{0}(\omega,\omega')=J^{0}(\omega,\omega'),
\label{(2.17)}
\end{equation}
and 
\begin{equation}
\Bigr(\Box^{0}-\xi R^{0}\Bigr)G^{1}(\omega,\omega')=J^{1}(\omega,\omega').
\label{(2.18)}
\end{equation}
With the notation in appendix B, one has
\begin{equation}
\Box^{0}=-{\partial^{2}\over \partial t^{2}}
-a_{1}(t){\partial \over \partial t}
-a_{2}(t)\bigtriangleup_{3},
\label{(2.19)}
\end{equation}
\begin{equation}
R^{0}={12\over L^{2}},
\label{(2.20)}
\end{equation}
\begin{equation}
J^{0}(\omega,\omega') = -{\delta(\omega-\omega') \over L^{3}\cosh^{3}(t/L)
\sin^{2}\theta_{1} \sin \theta_{2} },
\label{(2.21)}
\end{equation}
where $\bigtriangleup_{3}$ denotes the Laplacian on the 3-sphere, i.e.
\begin{equation}
\bigtriangleup_{3}=-\left[
{\partial^{2}\over \partial \theta_{1}^{2}}
+{2\over \tan \theta_{1}}{\partial \over \partial \theta_{1}}
+{1\over \sin^{2}\theta_{1}}{\partial^{2}\over \partial \theta_{2}^{2}}
+{\cot \theta_{2}\over \sin^{2}\theta_{1}}
{\partial \over \partial \theta_{2}}
+{1\over \sin^{2}\theta_{1} \sin^{2}\theta_{2}}
{\partial^{2}\over \partial \theta_{3}^{2}} \right],
\label{(2.22)}
\end{equation}
while the source-like term takes the form
\begin{equation}
J^{1}(\omega,\omega')={\cos \theta_{3}\over 2aL^{2}\sin \theta_{1}}\delta(\omega-\omega')
-(\Box^{1}-\xi R^{1})G^{0}(\omega,\omega'),
\label{(2.23)}
\end{equation}
having set (see again appendix B for all coefficient functions)
\begin{equation}
\Box^{1}=b_{1}(\omega){\partial^{2}\over \partial t^{2}}
+b_{2}(\omega){\partial \over \partial t}
+b_{3}(\omega){\partial \over \partial \theta_{1}}
+b_{4}(\omega){\partial \over \partial \theta_{2}} 
+b_{5}(\omega){\partial \over \partial \theta_{3}},
\label{(2.24)}
\end{equation}
and
\begin{equation}
R^{1}={3\over 2aL}\cosh(t/L)(7 \cosh(2t/L)-5)
\sin \theta_{1} \sin \theta_{2} \cos \theta_{3}.
\label{(2.25)}
\end{equation}
The zeroth-order Green function $G^{0}(\omega,\omega')$ has been already obtained
in the literature \cite{Bunch, Cher, Allen, Kirsten} and it reads as
\begin{equation}
G^{0}(\omega,\omega')={2\over (4\pi L)^{2}}\Gamma \left({3\over 2}-\lambda
\right) \Gamma \left({3\over 2}+\lambda \right)
F \left({3\over 2}-\lambda,{3\over 2}+\lambda,2;{1+Z \over 2}\right),
\label{(2.26)}
\end{equation}
where $\Gamma$ and $F$ are the standard notations for Gamma and 
hypergeometric function, respectively, while
\begin{equation}
\lambda^{2} \equiv {9\over 4}-12 \xi,
\label{(2.27)}
\end{equation}
and
\begin{equation}
Z \equiv {\cos \gamma -(\cos \eta)(\cos \eta') \over 
(\sin \eta)(\sin \eta')},
\label{(2.28)}
\end{equation}
where $\eta$ is the conformal time defined by
\begin{equation}
\eta \equiv 2 \arctan \; {\rm e}^{t/L},
\label{(2.29)}
\end{equation}
and $\gamma$ is the angle between $(\theta_{1},\theta_{2},\theta_{3})$
and $(\theta_{1}',\theta_{2}',\theta_{3}')$.
At this stage, the explicit evaluation of $G^{1}(\omega,\omega')$ remains rather difficult, which is
why we are also considering the broader framework outlined in the following section.

\section{Integral representation of the Green function}

Whenever one deals with partial differential operators on curved
Riemannian manifolds (for gauge theories, they can be viewed as acting 
on smooth sections of vector bundles over spacetime), the momentum-space
representation is no longer possible in general, because in a generic
curved spacetime the homogeneity properties needed for the use
of Fourier-transform techniques are lacking. Thus, the analysis of 
ultraviolet or infrared regimes for propagators becomes much harder.

However, mathematicians have developed a powerful symbolic calculus 
for pseudodifferential operators, which admit the differential
operators of theoretical physics as a particular case (but are non-local,
unlike the genuinely differential operators). The work of
Widom \cite{Widom} and Gusynin \cite{Gusynin}, among the others,
implies that, for a generic curved background in four dimensions,
the Green function $G(x,x')$ of the wave operator can be evaluated
according to the recipe
\begin{equation}
G(x,x')=(2\pi)^{-4}\lim_{\lambda \to 0} 
\int {{\rm d}^{4}k \over \sqrt{-\text{det} g_{\mu\nu}(x')}}
{\rm e}^{{\rm i}l(x,x',k)}\sigma(x,x',k;\lambda),
\label{(3.1)}
\end{equation}
where $l(x,x',k)$ is a {\it phase function} which reduces to the
familiar $k_{\mu}(x-x')^{\mu}$ in Minkowski spacetime, while
$\sigma(x,x',k;\lambda)$ is the corresponding {\it amplitude}.
The amplitude and phase functions make it possible to achieve a
geometric, manifestly covariant expression of the Green function
in coordinate space, despite the lack of the standard Fourier-transform
methods of flat-space field theory. 

In the integrand of (3.1), we now consider the following asymptotic expansions in the
neighbourhood of $\varepsilon=0$:
\begin{equation}
{1 \over \sqrt{- {\rm det} g_{\mu \nu}(x')}} \sim {1 \over \sqrt{-g_{0}(x')}}
+\varepsilon P_{1}(x')+{\rm O}(\varepsilon^{2}),
\label{(3.2)}
\end{equation}
\begin{equation}
l(x,x',k) \sim l_{0}(x,x',k) + \varepsilon l_{1}(x,x',k)+{\rm O}(\varepsilon^{2}),
\label{(3.3)}
\end{equation}
\begin{equation}
\sigma(x,x',k;\lambda) \sim \sigma_{0}(x,x',k;\lambda)+\varepsilon \sigma_{1}(x,x',k;\lambda)
+{\rm O}(\varepsilon^{2}).
\label{(3.4)}
\end{equation}
With our notation, $g_{0}$ is the determinant of the unperturbed metric (2.3) and $P_{1}$ is the
first-order perturbation which, in the closed-slicing coordinates, is the coefficient of
$\delta(\omega-\omega')$ in (2.23). Thus, writing hereafter $\omega_{1} \equiv t, \omega_{2}
\equiv \theta_{1},\omega_{3} \equiv \theta_{2},\omega_{4} \equiv \theta_{3}$, and by making a
comparison with (2.16), we find the following forms of $G^{0}$ and $G^{1}$ (since the asymptotic
expansions, if they exist, are unique):
\begin{equation}
G^{0}(\omega,\omega')=(2\pi)^{-4}\lim_{\lambda \to 0} \int {{\rm d}^{4}k \over \sqrt{-g_{0}(\omega')}}
{\rm e}^{{\rm i}l_{0}(\omega,\omega',k)}\sigma_{0}(\omega,\omega',k;\lambda),
\label{(3.5)}
\end{equation}
\begin{eqnarray}
G^{1}(\omega,\omega')&=& (2\pi)^{-4}\lim_{\lambda \to 0} \int {\rm d}^{4}k \; 
{\rm e}^{{\rm i}l_{0}(\omega,\omega',k)}\left[\sigma_{0}(\omega,\omega',k;\lambda)P_{1}(\omega')\right .
\nonumber \\
&+& \left. {\sigma_{1}(\omega,\omega',k;\lambda)
+{\rm i}l_{1}(\omega,\omega',k)\sigma_{0}(\omega,\omega',k;\lambda) \over
\sqrt{-g_{0}(\omega')}}\right].
\label{(3.6)}
\end{eqnarray}
At this stage, since the operator $\Box^{0}$ in Eq. (2.19) can be re-expressed as
\begin{equation}
\Box^{0}=\sum_{j=1}^{4}C_{j}(\omega){\partial^{2}\over \partial \omega_{j}^{2}}
+\sum_{j=1}^{3}D_{j}(\omega){\partial \over \partial \omega_{j}},
\label{(3.7)}
\end{equation}
where
\begin{equation}
C_{1}(\omega) \equiv -1,
\label{(3.8)}
\end{equation}
\begin{equation}
C_{2}(\omega) \equiv a_{2}(\omega_{1}),
\label{(3.9)}
\end{equation}
\begin{equation}
C_{3}(\omega) \equiv {a_{2}(\omega_{1}) \over \sin^{2}\omega_{2}},
\label{(3.10)}
\end{equation}
\begin{equation}
C_{4}(\omega) \equiv {a_{2}(\omega_{1}) \over (\sin^{2}\omega_{2}) (\sin^{2}\omega_{3})}, 
\label{(3.11)}
\end{equation}
\begin{equation}
D_{1}(\omega) \equiv -a_{1}(\omega_{1}), 
\label{(3.12)}
\end{equation}
\begin{equation}
D_{2}(\omega) \equiv {2 a_{2}(\omega_{1})\over \tan \omega_{2}},
\label{(3.13)}
\end{equation}
\begin{equation}
D_{3}(\omega) \equiv a_{2}(\omega_{1}) {\cot \omega_{3}\over \sin^{2}\omega_{2}},
\label{(3.14)}
\end{equation}
we find, upon defining
\begin{equation}
v_{0}(\omega,\omega',k) \equiv P_{1}(\omega') 
+{\rm i}{l_{1}(\omega,\omega',k)\over \sqrt{-g_{0}(\omega')}},
\label{(3.15)}
\end{equation}
\begin{equation}
v_{1}(\omega,\omega',k)=v_{1}(\omega')={1 \over \sqrt{-g_{0}(\omega')}},
\label{(3.16)}
\end{equation}
\begin{equation}
h_{j}(\omega,\omega',k;\lambda) \equiv {\partial \over \partial \omega_{j}} 
\Bigr[{\rm e}^{{\rm i}l_{0}}(v_{0}\sigma_{0}+v_{1}\sigma_{1})\Bigr],
\label{(3.17)}
\end{equation}
\begin{equation}
f_{j}(\omega,\omega',k;\lambda) \equiv {\partial \over \partial \omega_{j}}
h_{j}(\omega,\omega',k;\lambda),
\label{(3.18)}
\end{equation}
the following form of the partial differential equation for $G^{1}(\omega,\omega')$:
\begin{eqnarray}
(\Box^{0}-\xi R^{0})G^{1}(\omega,\omega')&=&
(2\pi)^{-4} \lim_{\lambda \to 0} \int {\rm d}^{4}k
\biggr[\sum_{j=1}^{4}C_{j}(\omega) f_{j}(\omega,\omega',k;\lambda) 
\nonumber \\
&+& \sum_{j=1}^{3}D_{j}(\omega)h_{j}(\omega,\omega',k;\lambda)
-{12 \xi \over L^{2}} 
{\rm e}^{{\rm i}l_{0}(\omega,\omega',k)} 
(v_{0}\sigma_{0}+v_{1}\sigma_{1})\biggr]
\nonumber \\
&=& J^{1}(\omega,\omega').
\label{(3.19)}
\end{eqnarray}
This equation should be solved for $l_{1}(\omega,\omega',k)$ and $\sigma_{1}(\omega,\omega',k;\lambda)$,
and our current work is aimed at finding the explicit solution.

\section{Concluding remarks}

The attempt of finding by direct calculation the Green function for a Casimir cavity in 
curved spacetime depends on the researcher's ability to solve an involved set of partial
differential equations in such a framework. From the mathematical point of view, this means
having the opportunity to learn a lot more on the solvability of hyperbolic equations on a manifold.
From the physical point of view, this can teach us new profound lessons on the physical implications
and applications of vacuum energy \cite{Call14}. More precisely, the theoretical work in \cite{Milt14}
has proved that Casimir energy gravitates like any other form of energy, and this holds 
independently of the orientation of the Casimir apparatus relative to the gravitational field.
The authors of \cite{Milt14} find that the total Casimir energy, including the divergent parts
which renormalize the masses of plates, possesses the gravitational mass demanded by the equivalence
principle (see also \cite{Saha04}). When the surface energy density residing 
on the Casimir plates is included, the integrated energy density equals the total energy.
It would be extremely interesting to understand to which extent such results can be extended
to the Casimir apparatus considered in our paper and in \cite{Chen12}).

We should also bring to the attention of the general reader the important work on the Casimir effect
in \cite{Eli10,Saha11}, where the authors have studied a massive scalar field 
with arbitrary curvature-coupling parameter, in the region between two infinite parallel plates,
on a de Sitter background. Among the many interesting results obtained therein, the decay of the
Casimir force at large plate separation has been shown to be power law, regardless of what value
the field mass takes.

\acknowledgments 
G. E. is grateful to the Dipartimento di Fisica of
Federico II University, Naples, for hospitality and support, and to
Enrico Calloni for more than a decade of joint work on Casimir cavities in a gravitational 
field. This short review paper
is dedicated to Volodya and Margarita Man'ko on the occasion of their 75th birthday.
Their full commitment to research and their brilliant work are a model for us and for
the generations to come. 

\appendix

\section{Christoffel symbols}
The explicit expression of the Christoffel symbols in closed-slicing coordinate system reads as follows. As above, 
by expanding in powers of $\varepsilon$ such connection coefficients, that is,
\begin{equation}
\Gamma_{\mu\nu}^\rho \sim { }^{(0)}\Gamma_{\mu\nu}^{\rho} 
+ \varepsilon { }^{(1)}\Gamma_{\mu\nu}^{\rho} + {\rm O}(\varepsilon^{2}),
\end{equation}
we find that the only non-vanishing terms are
\begin{gather}
{ }^{(0)}\Gamma_{22}^{1} =  \frac{L}{2} \sinh(2t/L), \\
{ }^{(0)}\Gamma_{33}^{1} =  \frac{L}{2} \sinh(2t/L) \sin^2\theta_1 , \\
{ }^{(0)}\Gamma_{33}^{2} = - \sin\theta_1 \cos\theta_1 , \\
{ }^{(0)}\Gamma_{44}^{1} =  \frac{L}{2} \sinh(2t/L) \sin^2\theta_1 \sin^2\theta_2  ,\\
{ }^{(0)}\Gamma_{44}^{2} = - \sin\theta_1 \cos\theta_1 \sin^2\theta_2 , \\
{ }^{(0)}\Gamma_{44}^{3} = - \sin\theta_2 \cos\theta_2 , \\
{ }^{(0)}\Gamma_{12}^{2} = { }^{(0)}\Gamma_{13}^{2} 
= { }^{(0)}\Gamma_{14}^{2} = \frac{L}{2} \tanh(t/L), \\
{ }^{(0)}\Gamma_{23}^{3} = { }^{(0)}\Gamma_{24}^{4} = \cot \theta_1 , \\
{ }^{(0)}\Gamma_{34}^{4} = \cot \theta_2 ,
\end{gather}
and
\begin{gather}
{ }^{(1)}\Gamma_{11}^{1} = \frac{3}{2 a}  \sinh (t/L) \cosh^2(t/L) \sin\theta_1 \sin\theta_2 \cos\theta_3 , \\
{ }^{(1)}\Gamma_{11}^{2}= \frac{1}{2 a L}  \cosh (t/L) \cos\theta_1  \sin\theta_2  \cos\theta_3 , \\
{ }^{(1)}\Gamma_{11}^{3} = \frac{1}{2 a L}  \cosh (t/L) \frac{\cos\theta_2 \cos\theta_3}  {\sin\theta_1} , \\
{ }^{(1)}\Gamma_{11}^{4} = -\frac{1}{2 a L}  \cosh (t/L) \frac{\sin\theta_3}{\sin\theta_1 \sin\theta_2}  , \\
{ }^{(1)}\Gamma_{12}^{1} = \frac{L}{2 a}  \cosh^3(t/L) \cos\theta_1 \sin\theta_2 \cos\theta_3 , \\
{ }^{(1)}\Gamma_{22}^{1} = -\frac{L^2}{a}  \sinh (t/L) \cosh^4(t/L)  \sin \theta_1 \sin\theta_2  \cos\theta_3 , \\
{ }^{(1)}\Gamma_{13}^{1} = \frac{L}{2 a}  \cosh^3(t/L) \sin\theta_1 \cos\theta_2 \cos\theta_3 , \\
{ }^{(1)}\Gamma_{33}^{1} = -\frac{L^2}{a}  \sinh (t/L) \cosh^4(t/L) \sin\theta_2 \cos\theta_3  \sin^3\theta_1  , \\
{ }^{(1)}\Gamma_{14}^{1} = -\frac{L}{2 a}  \cosh^3(t/L) \sin \theta_1 \sin \theta_2 \sin \theta_3 , \\
{ }^{(1)}\Gamma_{44}^{1} = -\frac{L^2 }{a}  \sinh (t/L) \cosh^4(t/L)  \sin^3\theta_1 \sin^3\theta_2 \cos\theta_3 .
\end{gather}

\section{Notation for our differential operators}

In section 2 the coefficients occurring in our second-order partial differential operators 
read as follows:
\begin{equation}
a_{1}(t) \equiv {3\over L}\tanh (t/L),
\end{equation}
\begin{equation}
a_{2}(t) \equiv {1\over L^{2} \cosh^{2}(t/L)},
\end{equation}
\begin{equation}
b_{1}(\omega) \equiv {L \over a}\cosh^{3}(t/L)(\sin \theta_{1})
(\sin \theta_{2}) (\cos \theta_{3}),
\end{equation}
\begin{equation}
b_{2}(\omega) \equiv {9\over 2a}\cosh^{2}(t/L)\sinh(t/L)
(\sin \theta_{1})(\sin \theta_{2})(\cos \theta_{3}),
\end{equation}
\begin{equation}
b_{3}(\omega) \equiv {1\over 2aL}\cosh(t/L)(\cos \theta_{1})
(\sin \theta_{2})(\cos \theta_{3}),
\end{equation}
\begin{equation}
b_{4}(\omega) \equiv {1\over 2aL}\cosh(t/L)
{(\cos \theta_{2})(\cos \theta_{3})\over \sin \theta_{1}},
\end{equation}
\begin{equation}
b_{5}(\omega) \equiv {1\over 2aL}\cosh(t/L)
{\sin \theta_{3} \over (\sin \theta_{1})(\sin \theta_{2})}.
\end{equation}

\end{document}